\documentclass[aps,pra,reprint,showpacs,superscriptaddress,10pt]{revtex4-1}
\usepackage{amsmath,amssymb,amsfonts,graphicx,bm,longtable,dcolumn,color,textcomp,lmodern,mathrsfs}
\usepackage[utf8]{inputenc}
\usepackage[T1]{fontenc}

\begin{document}

\title{Dynamical response of a radiative thermal transistor based on suspended insulator-metal transition membranes}

\author{Ivan Latella}
\affiliation{Department of Mechanical Engineering, Universit\'{e} de Sherbrooke, Sherbrooke, PQ J1K 2R1, Canada}

\author{Olivier Marconot}
\affiliation{Department of Mechanical Engineering, Universit\'{e} de Sherbrooke, Sherbrooke, PQ J1K 2R1, Canada}

\author{Luc Fr\'echette}
\affiliation{Department of Mechanical Engineering, Universit\'{e} de Sherbrooke, Sherbrooke, PQ J1K 2R1, Canada}

\author{Julien Sylvestre}
\affiliation{Department of Mechanical Engineering, Universit\'{e} de Sherbrooke, Sherbrooke, PQ J1K 2R1, Canada}

\author{Philippe Ben-Abdallah}
\email{pba@institutoptique.fr}
\affiliation{Department of Mechanical Engineering, Universit\'{e} de Sherbrooke, Sherbrooke, PQ J1K 2R1, Canada}
\affiliation{Laboratoire Charles Fabry, UMR 8501, Institut d'Optique, CNRS, Universit\'{e} Paris-Saclay, 2 Avenue Augustin Fresnel, 91127 Palaiseau Cedex, France}

\begin{abstract}
We investigate the dynamical control of the heat flux exchanged in near-field regime between a membrane made with a phase-change material and a substrate when the temperature of the membrane is tuned around its critical value. We show that in interaction with an external source of thermal radiation, this system is multistable and behaves as a thermal transistor, being able to dynamically modulate and even amplify super-Planckian heat fluxes. This behavior could be used to dynamically control heat fluxes exchanged at the nanoscale in systems out of thermal equilibrium and to process thermal information employing suspended membranes. 
\end{abstract}

\maketitle

\section{Introduction}

Controlling heat exchanges at the nanoscale is a challenging problem both from a fundamental point of view and for the development of new technologies.  While it is nowadays common to control electric currents since the invention of the electronic solid-state transistor~\cite{Bardeen}, it is unfortunately much less common to exercise the same control over heat. 

In 2006, Li et al.~\cite{Casati1} introduced a thermal counterpart of field-effect transistor to control heat fluxes carried by phonons through solid segments, paving the way for building blocks to process information~\cite{BaowenLiEtAl2012,BaowenLi2} using heat flux rather than electric currents. More recently, the concept of transistor has been extended to contactless out of thermal equilibrium systems~\cite{PBA_PRL2014}.  In this case, heat flows with the passage of thermal photons from one material to another. 

The radiative thermal transistor consists of three elements called, by analogy with its electronic counterpart, source, drain and gate. The source and drain are made with solids held at different temperatures to create a temperature gradient. The source, being traditionally hotter than the drain, emits thermal photons which transfer heat to the drain. These two solids are separated by an intermediate layer made of an insulator-metal transistion (IMT) material~\cite{PBA_APL}. This layer plays the role of the gate. By tuning the gate temperature around its critical value, it is possible to drastically change the flux received by the drain and even to amplify this flux. The device can work either at large separation distances (far-field regime~\cite{Joulain2015}), where heat fluxes are associated to propagating photons, or at short distances (near-field regime~\cite{PBA_PRL2014}), where heat is transferred mainly by photon tunneling. Besides modulation and amplification of heat fluxes, these structures based on phase-change materials can be used to store thermal energy~\cite{Kubytskyi,ItoEtAl2016} and to make logical operations~\cite{PBAlogic} with thermal photons.

One of the main advantages of a near-field thermal transistor, beyond its compactness, is its capability to manipulate super-Planckian heat fluxes~\cite{Polder,HargreavesPLA69,KittelPRL05,NarayanaswamyPRB08,HuApplPhysLett08,ShenNanoLetters09,RousseauNaturePhoton09,OttensPRL11,KralikRevSciInstrum11,KralikPRL12,SongNatureNano15,KimNature15,StGelaisNatureNano16,KloppstecharXiv,WatjenAPL16,Guha,Tschikin}, that is, heat fluxes which are larger than the fluxes radiated by blackbodies at the same temperature.  
Such a transistor is, however, a solid element with three terminals that must be positioned at close separation distances, a fact that currently limits the development of this technology. In the present work, we investigate the heat transfer between a thermal bath and a substrate that are separated by a thin membrane made of an IMT material. Different works have already shown~\cite{van_Zwol1,van_Zwol2,van_Zwol3} that the heat flux exchanged in the near field between an IMT material and another medium can be modulated by several orders of magnitude across the phase transition of this material.
Here the membrane is maintained above the substrate at a sub-micrometer separation. We show that this two-body system in the presence of the thermal bath exhibits the same features of a three-body transistor. 

Specifically, the radiative thermal transistor under investigation consists of a gate made of a membrane of vanadium dioxide (VO$_2$) at temperature $T_G$ which is suspended over a slab of silicon dioxide (SiO$_2$) at temperature $ T_D = 300\, $K acting as the drain. This structure is illuminated with thermal radiation coming from a far-field source assumed to be a blackbody at temperature $ T_S $. 
At a certain critical temperature, the VO$_2$ membrane undergoes an insulator-to-metal transition in which the insulating and metallic phases coexist over a finite temperature range~\cite{Mott}. 
We describe the transition region with $ T_c $ and $ \Delta $, in such a way that it takes place in the temperature range between $ T_c $ and $ T_c + \Delta $. Because of hysteresis, when increasing and decreasing the temperature of the gate, a shift in the temperature dependence of the VO$_2$ properties is observed and has been reported to be about $ 8\, $K~\cite{Mott} in the studied situation. According to this, we take the critical temperature as $ T_c = 341\, $K for increasing $ T_G $ and $ T_c = 333\, $K when $ T_G $ is decreased, while the width of the transition is $ \Delta = 4\, $K in both cases.  

The gate and the drain are separated by a subwavelength distance $ d $ (at room temperature, the thermal wavelength is about 8\,$\mu$m), 
so that they are coupled in the near-field regime. In this regime, in addition to the usual propagating waves, evanescent waves also contribute to the heat transfer. Moreover, in the insulating phase, VO$_2$ supports surface waves that interact with the surface modes supported by SiO$_2$, leading to an efficient energy exchange. In the metallic phase, however, VO$_2$ does not support resonant modes and thus, the heat exchange with the drain is much less efficient. Such a behavior induces a negative differential thermal resistance, since when the temperature of the gate is increased around the critical temperature, the heat exchange with the drain is decreased~\cite{PBA_PRL2014}. Furthermore, propagating photons coming from the source are more easily transmitted to the drain when VO$_2$ behaves as an insulator, because the screening of electromagnetic fields is stronger in the metallic phase.
Here, the thickness of the gate is set to $ \delta = 200\, $nm and the drain is assumed to be semi-infinite. 

\begin{figure}
\includegraphics[scale=0.37]{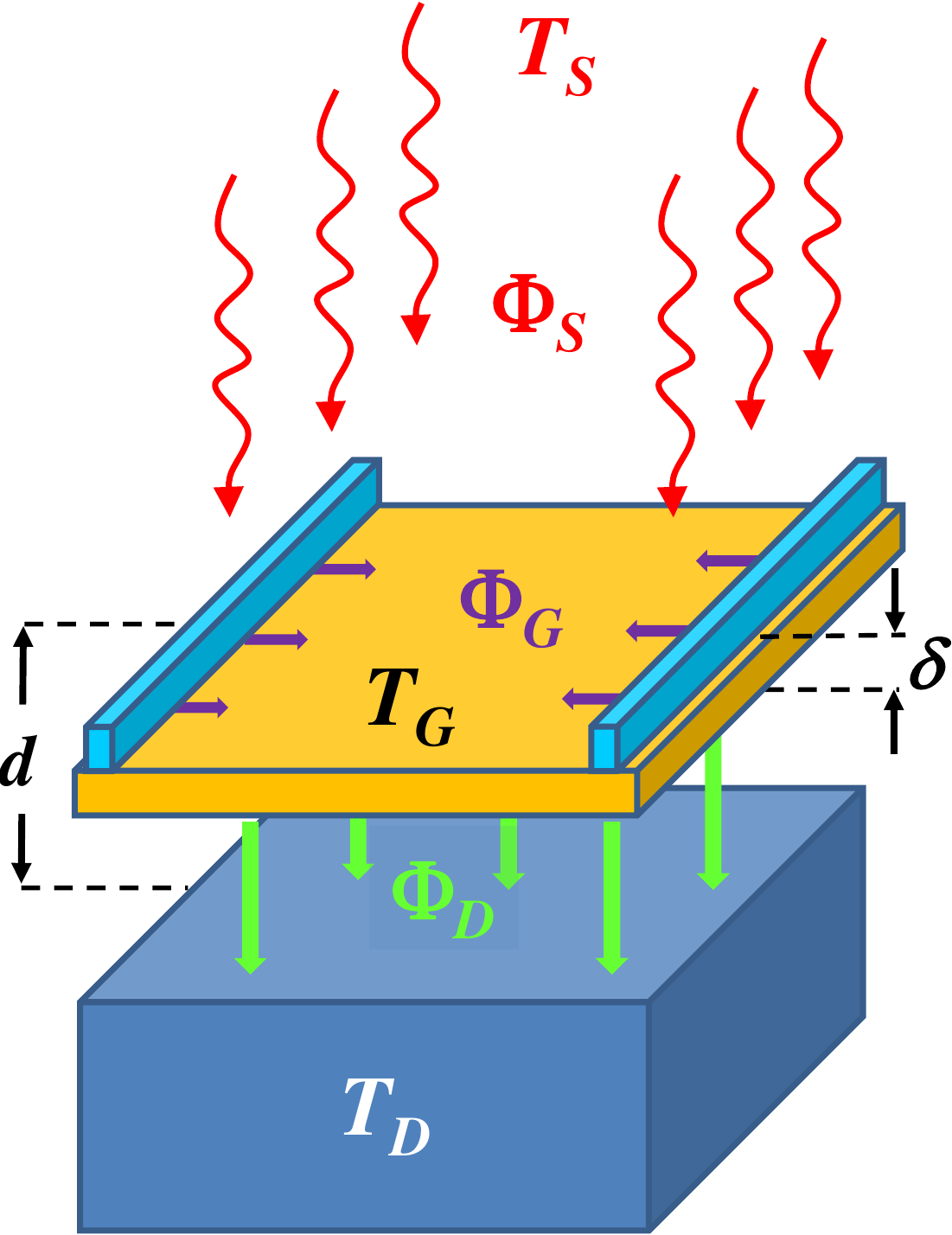}
\caption{Sketch of a radiative thermal transistor. A thermal source in the far field, here modeled as a blackbody at temperature $ T_S $, radiates towards the gate-drain structure. The gate is a membrane of VO$_2$ (IMT material) at temperature $ T_G $ suspended over a slab of SiO$_2$ at temperature $ T_D $ which acts as the drain. The gate and the drain are separated by the subwavelength distance $ d $. The energy fluxes radiated by the source and absorbed by the drain are denoted by $ \Phi_S $ and $ \Phi_D $, respectively, while $ \Phi_G $ is an external energy flux supplied to the gate. Energy can be added to the gate by heating or can be removed from it by cooling, for instance, using Peltier elements which are represented on top of the gate.}
\label{fig1}
\end{figure}

The proposed device is depicted in Fig.~\ref{fig1}. As we will show below, the interaction of the gate and the drain in the near field, which can be tuned by adjusting the separation distance $ d $, offers high versatility in the design of a device suitable for managing thermal fields radiated from far-field sources. 
After describing the radiative heat exchanges and showing how to calculate the equilibrium temperatures in this system, we investigate its dynamical evolution under the action of a temporal control of the gate temperature.

\section{Heat fluxes and equilibrium temperatures}
\label{sec:stationary_regime}

Given a thermal source, the radiative flux of energy arriving to the drain depends on the state of the gate which can be externally manipulated. The net energy flux on the gate is given by
\begin{equation}
\Phi = \Phi_S - \Phi_D + \Phi_G,
\label{flux}
\end{equation}
where $ \Phi_S $ and $ \Phi_D $ are the averaged component normal to the surfaces of the Pointyng vector in the regions between the source and the gate and between the gate and the drain, respectively, and $ \Phi_G $ is the energy flux externally supplied to the gate that controls the state of the device.
When $ \Phi_G > 0 $, an external energy flux is added to the gate by heating. When $ \Phi_G < 0 $, energy is removed from the gate by cooling, for instance, using Peltier elements.
In stationary states, the flux $ \Phi $ on the membrane vanishes and therefore, the external flux supplied to the gate equals minus the net radiative contribution,
\begin{equation}
\Phi_G = - ( \Phi_S - \Phi_D ).
\label{stationary_state}
\end{equation}

\begin{figure}
\includegraphics[scale=0.7]{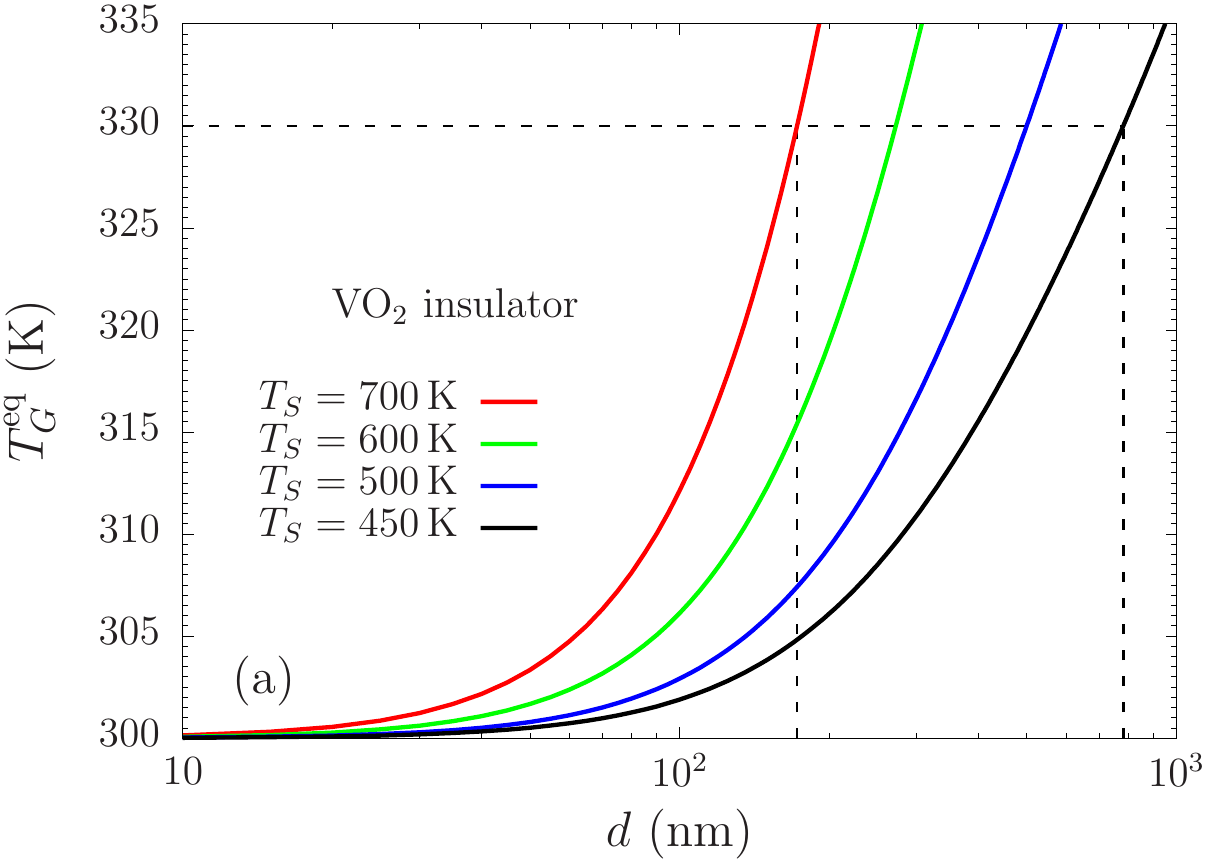}
\includegraphics[scale=0.7]{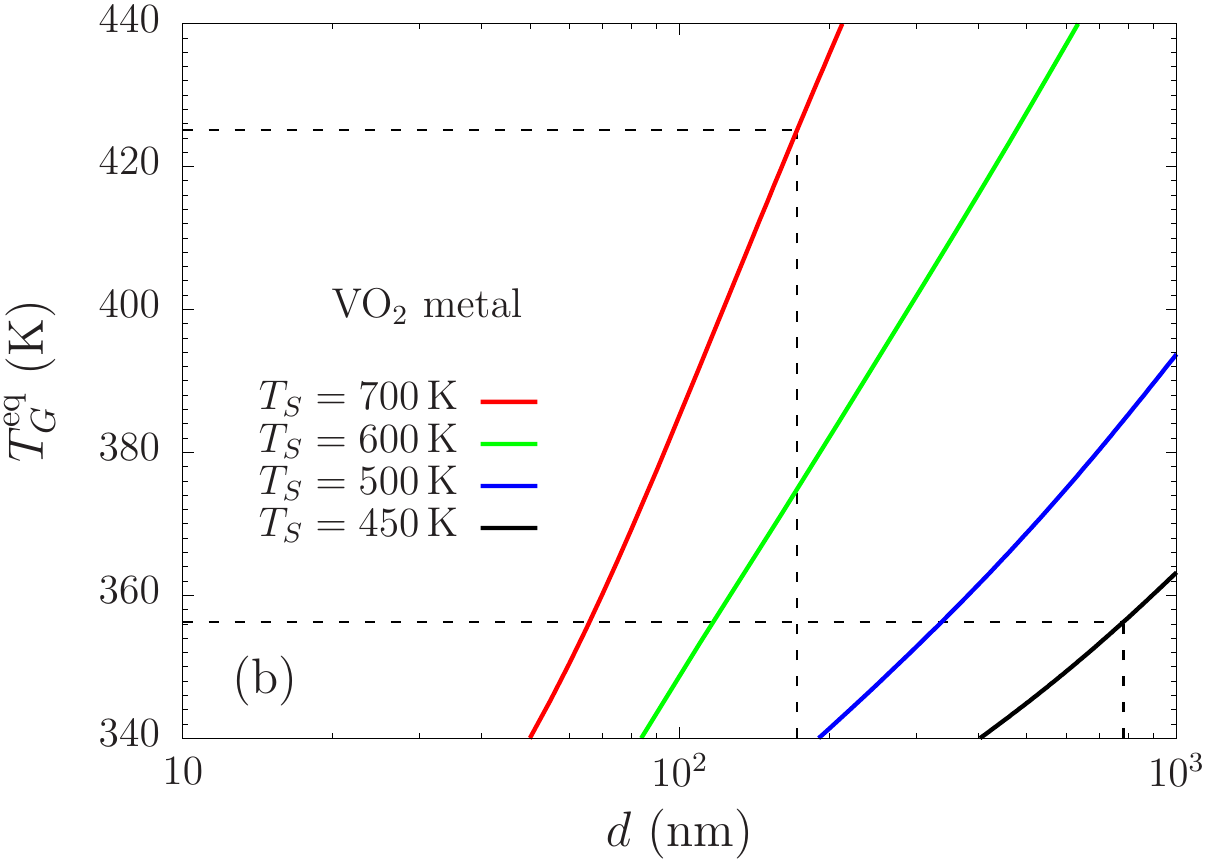}
\caption{Equilibrium temperatures of the gate $ T_G^\mathrm{ eq } $ as a function of the separation distance $ d $ for several source temperature $ T_S $. Here the thickness of the membrane is set to $ \delta = 200\, $nm and the drain temperature to $ T_D = 300\, $K. In (a), the membrane of VO$_2$ (the gate) is in the insulating phase and in (b), in the metallic phase. The dashed lines indicate particular choices of equilibrium temperatures and separation distances: In (a), $ T^\mathrm{ eq }_G = 330\, $K corresponds to the separation distances $ d = 172\, $nm and $ d = 782\, $nm for $ T_S = 700\, $K and $ T_S = 450\, $K, respectively, while in (b), these separation distances correspond to equilibrium temperatures $ T^\mathrm{ eq }_G = 425\, $K and $ T^\mathrm{ eq }_G = 356\, $K, respectively.}
\label{fig2}
\end{figure}

\begin{figure*}
\includegraphics[scale=0.7]{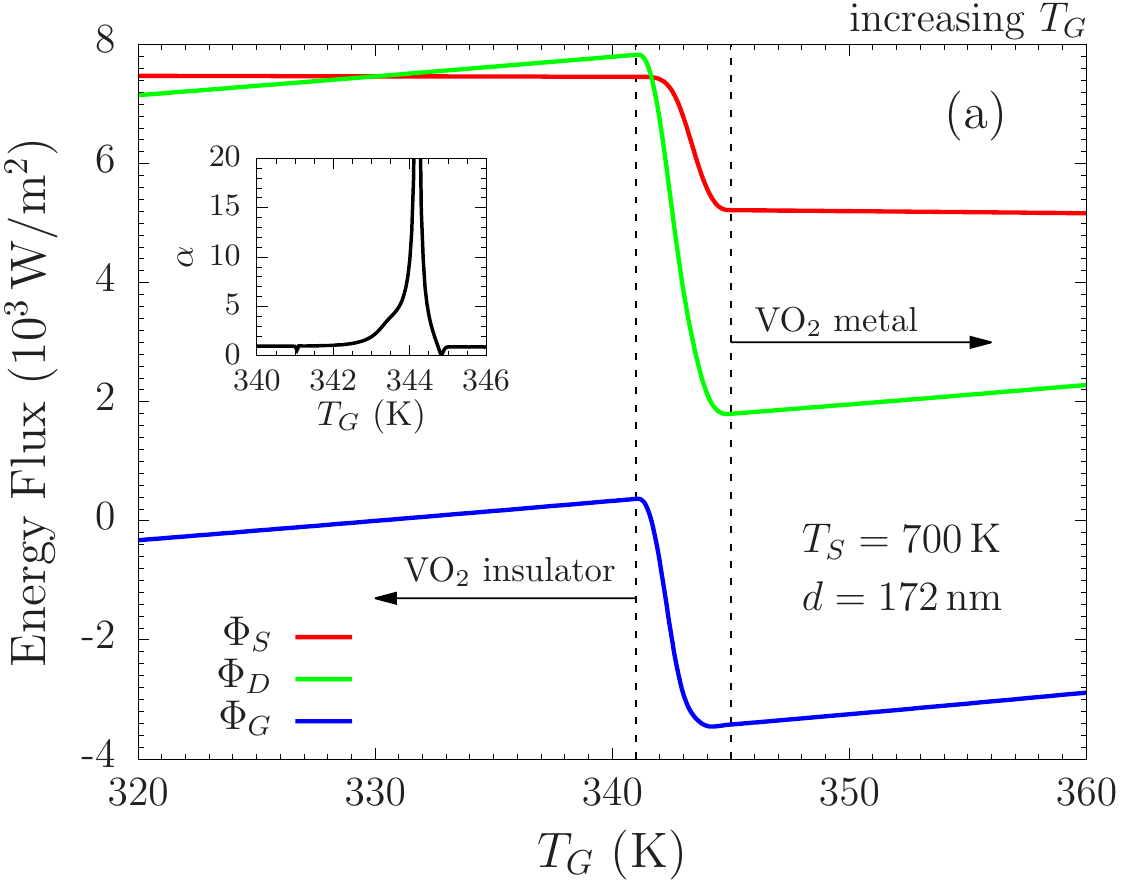}~\includegraphics[scale=0.7]{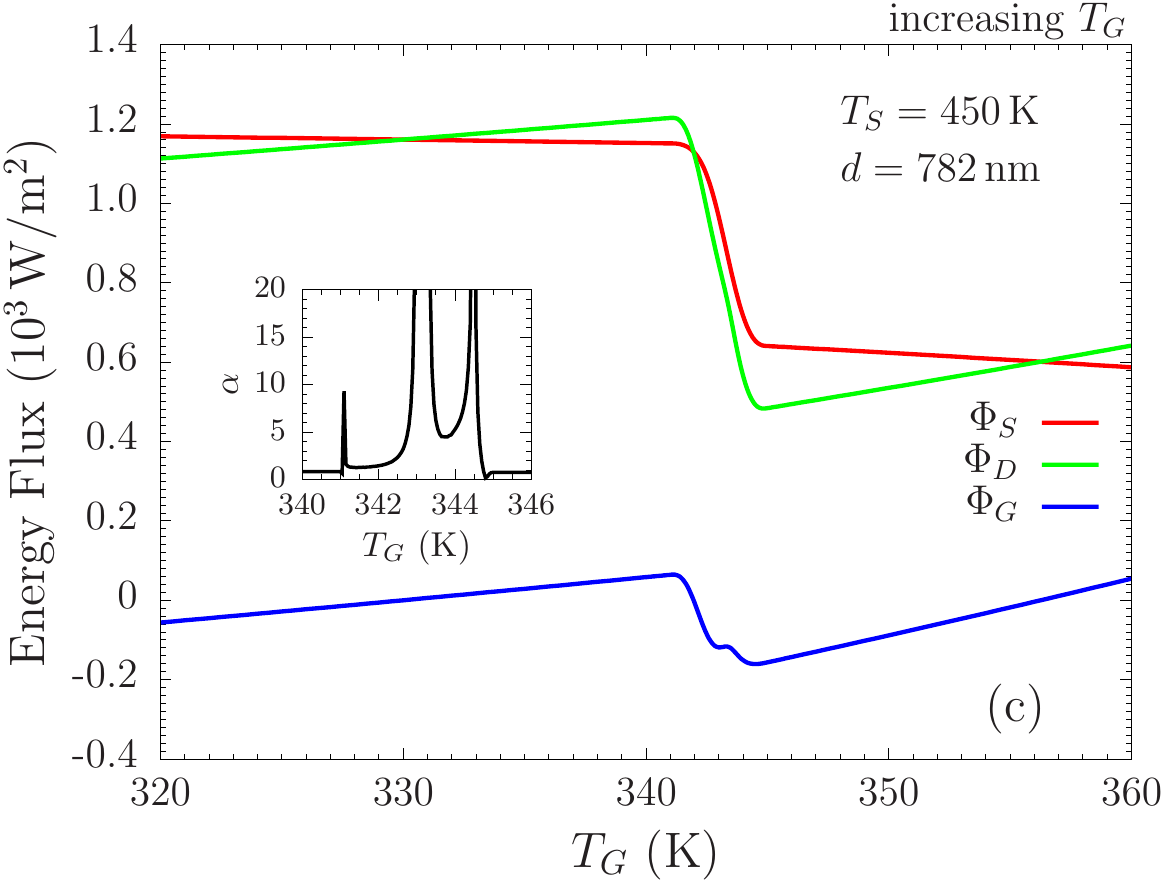}
\includegraphics[scale=0.7]{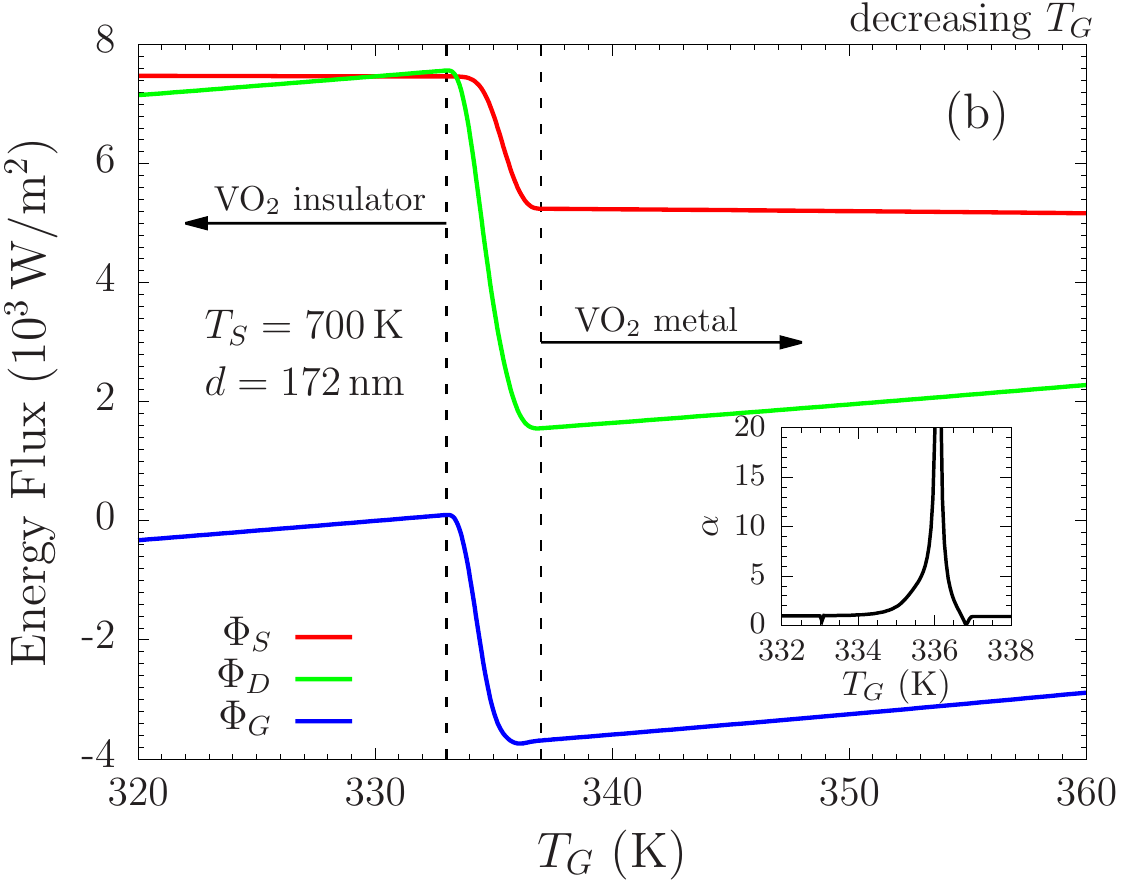}~\includegraphics[scale=0.7]{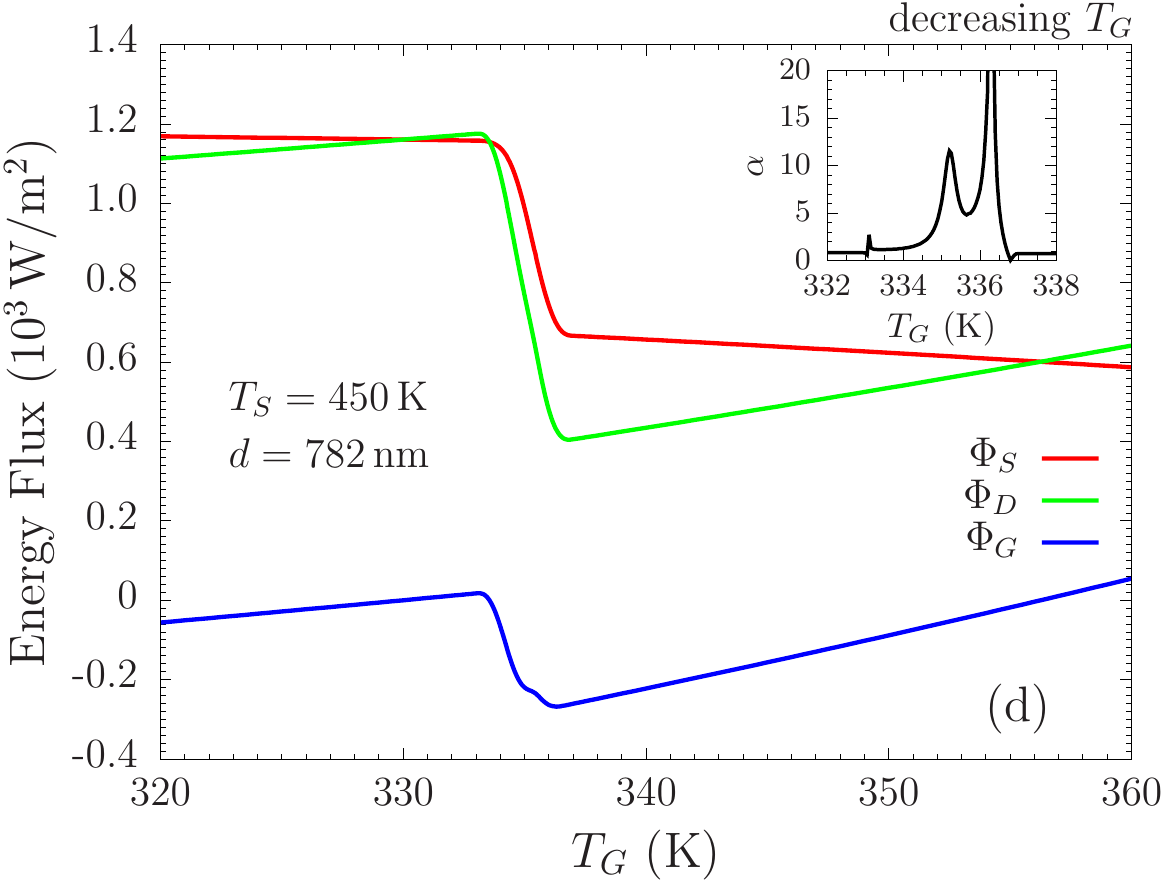}
\caption{Energy fluxes in the device. In (a) and (b) the fluxes are shown as a function of the gate temperature for increasing and decreasing $ T_G $, respectively, for $T_S=700\,$K and $d=172\,$nm.
In (c) and (d) the fluxes are shown when increasing and decreasing $ T_G $, respectively, for $T_S=450\,$K and $d=782\,$nm. The insets show the corresponding amplification factor.}
\label{fig3}
\end{figure*}

The energy flux carried by the electromagnetic field radiated in the regions of space of interest can be obtained using Rytov's fluctuational electrodynamics approach~\cite{Rytov,Joulain2005,Volokitin}, which assumes that the bodies radiate at local thermal equilibrium and that field correlations can be described using the fluctuation-dissipation theorem.
In this way, expanding the field in modes of frequency $ \omega $, wave vector parallel to the surfaces $ k $, and polarization $ p $, the radiative energy fluxes can be written as
\begin{align}
\Phi_S & = \int_0^\infty \frac{ d \omega }{ 2\pi } \hbar \omega \int_0^\infty \frac{ dk }{ 2\pi } k \sum_p
\Big( n_{ SG } \hat{ \mathcal{ T } }_S + n_{ GD } \hat{ \mathcal{ T } }_G \Big), \label{Phi_S} \\
\Phi_D & = \int_0^\infty \frac{ d \omega }{ 2\pi } \hbar \omega \int_0^\infty \frac{ dk }{ 2\pi } k \sum_p
\Big( n_{ SG } \hat{ \mathcal{ T } }_G + n_{ GD } \hat{ \mathcal{ T } }_D \Big), \label{Phi_D} 
\end{align}
where $ n_{ ij } = n_i - n_j $ is the difference of thermal photon distributions 
$ n_j ( \omega ) = 1 / \left( e^{ \hbar \omega / k_B T_j } - 1 \right) $ at temperature $ T_j $ with $ i,j = S,G,D $, $ k_B $ being the Boltzmann constant and $ \hbar $ the reduced Planck constant.
Here $ \hat{ \mathcal{ T } }_j = \hat{ \mathcal{ T } }_j ( \omega, k, p) $, $ j = S,G,D $, are the associated energy transmission coefficients given by~\cite{Latella2017}
\begin{align}
\hat{ \mathcal{ T } }_S & = 
\Pi^\mathrm{ pw } \big( 1 - | \rho_{ GD } |^2 \big),\\
\hat{ \mathcal{ T } }_G & = 
\Pi^\mathrm{ pw } \frac{ | \tau_G |^2  \big( 1 - | \rho_D |^2 \big) }{ \big| 1 - \rho_G \rho_D e^{ 2i k_z d } \big|^2 }, \\
\begin{split}
\hat{ \mathcal{ T } }_D & = 
\Pi^\mathrm{ pw } \frac{ \big( 1- | \rho_G |^2 \big) \big( 1 - | \rho_D |^2 \big) } 
{ \big| 1 - \rho_{ G } \rho_{ D } e^{ 2i k_z d } \big|^2 } \\
& + \Pi^\mathrm{ ew } \frac{ 4 \mathrm{ Im } ( \rho_G ) \mathrm{ Im } ( \rho_D ) e^{ -2 \mathrm{ Im }( k_z ) d } }
{ \big| 1 - \rho_G \rho_D e^{ 2i k_z d } \big|^2 },
\end{split}
\end{align}
where $ k_z = \sqrt{ \omega^2 / c^2 - k^2 } $ is the normal component of the wave vector ($ c $ being the speed of light in vacuum), $ \Pi^\mathrm{ pw } $ and $ \Pi^\mathrm{ ew } $ are projectors onto the propagating and evanescent wave sectors, respectively, $ \tau_G = \tau_G ( \omega, k, p ) $ is the optical transmission coefficient of the gate, $ \rho_G = \rho_G ( \omega, k, p ) $ and $\rho_D = \rho_ D ( \omega, k, p ) $ are the optical reflection coefficients of the gate and the drain, respectively, and $\rho_{ GD } = \rho_{ GD }( \omega, k, p ) $ is the reflection coefficient of the gate and the drain together considered as a single object. These optical reflection and transmission coefficients depend on the permittivity of the constituting materials: for SiO$_2$ we take the permittivity from Ref.~\cite{Palik} and from Ref.~\cite{Barker} for VO$_2$ using an effective medium theory~\cite{Biehs} in the transition region to account for the phase coexistence (see Sec.~\ref{sec:dynamics} for details on the description of the transition region).

As mentioned previously, tuning the near-field interaction between the gate and the drain by adjusting the separation distance allows us to set the device in appropriate working conditions. Of particular importance is the equilibrium temperature of the gate $ T^\mathrm{ eq }_G $ at fixed source and drain temperatures when no external energy flux is acting on the membrane. For practical purposes, we want to set this temperature to be $ T_G^\mathrm{ eq } = 330\, $K, in such a way that it remains right below the transition region. We will see that for a given $ T_S $, one can choose a suitable $ d $ leading to the desired gate equilibrium temperature.

From Eq.~(\ref{flux}), in the stationary regime and at zero applied flux, the equilibrium temperature is implicitly given by the relation
\begin{equation}
\Phi_S ( T_G^\mathrm{ eq }, T_S, T_D, d ) = \Phi_D ( T_G^\mathrm{ eq }, T_S, T_D, d ),
\label{condition_eqtemp}
\end{equation}
where $ \Phi_S $ and $ \Phi_D $ can be computed using Eqs.~(\ref{Phi_S}) and (\ref{Phi_D}), respectively.
If $ T_D = 300\, $K is held fixed, then Eq.~(\ref{condition_eqtemp}) can be solved to give the gate equilibrium temperature implicitly depending on the source temperature and the separation distance, $ T_G^\mathrm{ eq } = T_G^\mathrm{ eq } ( T_S, d ) $. 
Since the optical properties of VO$_2$ depend on temperature, however, the solution of Eq.~(\ref{condition_eqtemp}) may not be unique.
In Fig.~\ref{fig2}(a), we show $ T_G^\mathrm{ eq } $ as a function of $ d $ for several source temperatures $ T_S $ when VO$_2$ behaves as an insulator. The separation distances leading to $ T_G^\mathrm{ eq } = 330\, $K are indicated with dashed lines in this figure for two different source temperatures, namely, for $ T_S = 700\, $K we have to take $ d = 172\, $nm and for $ T_S = 450\, $K the separation has to be $ d = 782\, $nm.   These separation distances, in turn, correspond to equilibrium temperatures that are realizable when the membrane behaves as a metal. In Fig.~\ref{fig2}(b), we show $ T_G^\mathrm{ eq } $ for this case as a function of $ d $ for different $ T_S $. In the plot, with dashed lines we indicate the equilibrium temperatures corresponding to the given separations; $ T^\mathrm{ eq }_G = 425\, $K for $ d = 172\, $nm and  $ T^\mathrm{ eq }_G = 356\, $K for $ d = 782\, $nm. 
According to the precedent discussion, we have characterized the stationary states of the transistor at zero applied power by determining the gate temperatures associated with those states. Next we describe the energy fluxes in the system at stationary states with a nonvanishing external energy flux $ \Phi_G $.

We recall that the energy fluxes $ \Phi_S $ and $ \Phi_D $ are given by Eqs.~(\ref{Phi_S}) and (\ref{Phi_D}), respectively. In Fig.~\ref{fig3}(a), these fluxes are shown as a function of the gate temperature for the device configuration corresponding to the source at $ T_S = 700\, $K, in which $ d = 172\, $nm. There, we also show the energy flux applied to the gate $\Phi_G$ fulfilling the stationary state condition Eq.~(\ref{stationary_state}). In this figure we consider that the temperature of the VO$_2$ membrane is increased, while in Fig.~\ref{fig3}(b) the fluxes are shown in the case in which this temperature is decreased. The fluxes for the case in which $T_S=450\,$K and $d=782\,$nm are shown in Fig.~\ref{fig3}(c) and Fig.~\ref{fig3}(d) for increasing and decreasing $ T_G $, respectively.

Furthermore, the ability of the transistor to amplify energy fluxes can be described by the amplification factor~\cite{Casati1}
\begin{equation}
\alpha \equiv \left| \frac{ \partial \Phi_D }{ \partial \Phi_G } \right| 
= \left| 1 - \frac{ \Phi_S' }{ \Phi_D' } \right|^{ -1 },
\end{equation}
where in the second equality, we have used the stationary state condition Eq.~(\ref{stationary_state}), in which the primes denote derivative with respect to $ T_G $. Using an IMT material such as VO$_2$, as noted previously, induces a negative differential thermal resistance in the system, a necessary ingredient to obtain amplification factors larger than unity~\cite{Casati1,PBA_PRL2014}. The insets of Fig.~\ref{fig3} show the amplification factor corresponding to each situation, where large values of this quantity can be observed in the transition region. We note that $ \alpha $ here quantifies differential increments of the fluxes around a given stationary state.

\begin{figure}
\includegraphics[width=\columnwidth]{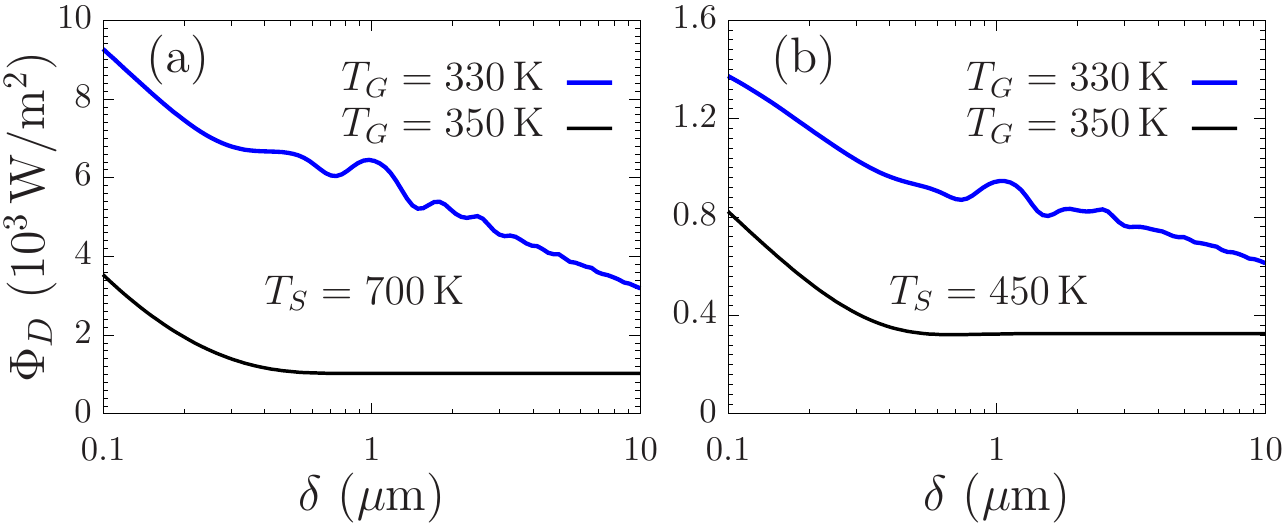}
\caption{Energy flux on the drain $\Phi_D$ as a function of the thickness $\delta$ of the VO$_2$ membrane.
In (a) the flux is shown for a source temperature $T_S=700\,$K and a separation distance $d=172\,$nm, while in (b) for $T_S=450\,$K and $d=782\,$nm. The flux is computed for two gate temperatures corresponding to the insulator ($T_G=330\,$K) and metallic ($T_G=350\,$K) phases of VO$_2$. In all cases, the temperature of the drain is set to $T_D = 300\,$K.}
\label{fig4}
\end{figure}

To conclude this section, we briefly consider the influence of the thickness of the membrane on the energy flux received by the drain. In Fig.~\ref{fig4}, we show the energy flux $ \Phi_D $ as a function of $ \delta $ for the fixed source temperatures and separation distances that we have used in our examples ($ T_S = 700\, $K and $ T_S = 450\, $K with the corresponding separation distances $ d = 172\, $nm and $ d = 782\, $nm, respectively). In addition, $ \Phi_D $ is shown for gate temperatures $ T_G = 330\, $K and $ T_G = 350\, $K at which the VO$_2$ membrane behaves as an insulator and as a metal, respectively. We observe that small variations of the thickness $ \delta $ around a given value do not induce large variations of the flux $ \Phi_D $.

After having analyzed the device in stationary conditions, in Sec.~\ref{sec:dynamics} we study the thermal relaxation of the gate and describe a dynamical modulation of the flux received by the drain.

\section{Dynamical modulation of heat flux received by the drain}
\label{sec:dynamics}

\begin{figure}
\includegraphics[scale=0.7]{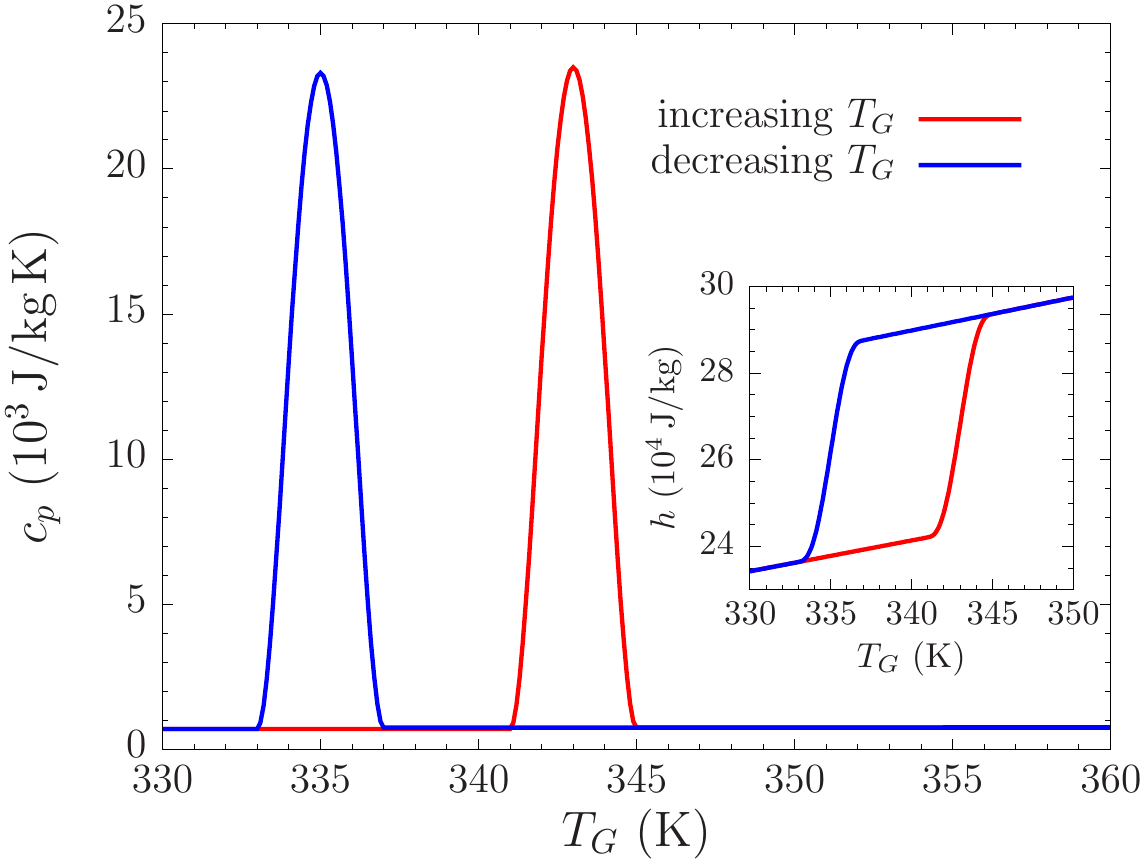}
\caption{Specific heat of the VO$_2$ membrane for increasing and decreasing $ T_G $. In the inset we show the associated enthalpy changes.}
\label{fig5}
\end{figure}

We are interested in studying the behavior of the transistor in situations out of the stationary state. 
Characterizing the dynamics of the device is relevant, for instance, for thermal information treatment, since transient behavior is inherent to the operation of transistors and its time response is an indication of the achievable processing rate.
Because the transistor is controlled by acting on the gate to modify its temperature, we first describe the behavior of the gate temperature in such nonequilibrium situations.

The behavior of the temperature of the gate $ T_G(t) $ as a function of the time $ t $ depends on the total energy flux on the gate $ \Phi $, given by Eq.~(\ref{flux}), and satisfies the evolution equation
\begin{equation}
\frac{ d T_G ( t ) }{ dt } = \frac{ \Phi ( T_G ) }{ I_G }, 
\label{evolution_temperature}
\end{equation}
where $ I_G = c_p \rho \delta $ is the thermal inertia of the gate, $ \rho = 4.67 \times 10^3\, $kg/m$^3$~\cite{Leroux} and $ c_p $ being the density and the specific heat per unit mass of VO$_2$, respectively. During the transition, $ c_p $ sharply increases and therefore the thermal inertia of the gate undergoes an abrupt change that modifies the temporal evolution of the membrane. Before evaluating the relaxation in time of the gate, we discuss a simple model to quantify the specific heat in the transition region. 

We recall that the transition region takes place for temperatures between $ T_c $ and $ T_c + \Delta $ (we assume $ T_c = 341\,$ K for increasing $ T_G $ and $ T_c = 333\, $K for decreasing $ T_G $, while $ \Delta = 4\, $K in both cases).
The state of the material in the transition region can be described by introducing a volume fraction $ f ( T ) $ such that $ f ( T_c ) = 0 $ and $ f ( T_c + \Delta ) = 1 $.
In real situations, this volume fraction depend on the actual working conditions of the device and can be estimated, for instance, by combining conductivity measurements in the transition region and an effective medium theory~\cite{Mott}.
Here, to keep our theoretical description as simple as possible, we model the volume fraction as a smooth step function given by 
\begin{equation}
f = 6 x^5 - 15 x^4 + 10 x^3, \qquad x = \frac{ T-T_c }{ \Delta },
\end{equation}
whose first and second derivatives vanish at the ends of the transition region. 
This volume fraction is used, for instance, to compute the radiative heat fluxes by means of an effective medium theory~\cite{Biehs}, quantifying the permittivity of VO$_2$ in the transition region.

The enthalpy per unit mass of VO$_2$ can be written as
\begin{equation}
h=
\begin{cases}
h_i, & T \leq T_c \\
h_i + ( h_m - h_i ) f, &  T_c < T < T_c + \Delta \\
h_m, & T \geq T_c + \Delta
\end{cases} ,
\label{enthalpy}
\end{equation}
where, assuming that the specific heat is approximately constant before and after the transition, the enthalpies in the insulating and metallic phases $h_i$ and $h_m$, respectively, can be written as
\begin{align}
h_i & = c_{ p,i } T, \label{enthalpy_i}\\
h_m & = c_{ p,m }( T -T_c -\Delta ) + c_{ p,i } T_c + L.
\label{enthalpy_m}
\end{align}
Here $ L $ is the latent heat of the transition and $ c_{ p,i } = 710\, $J/(kg\,K) and $ c_{ p,m } = 760\, $J/(kg\,K) are the specific heats in the insulating and metallic phases~\cite{Berglund}, respectively. We note that, because of the hysteresis, the enthalpy is different for increasing and decreasing $ T_G $, which can be accounted for in Eqs.~(\ref{enthalpy}), (\ref{enthalpy_i}), and (\ref{enthalpy_m}) by choosing the appropriate $ T_c $ and $ L $. When $ T_G $ is increased, we take the latent heat as $ L = 5.15 \times 10^4\, $J/kg~\cite{Berglund}, while the corresponding one for decreasing $ T_G $ can be deduced from this value by requiring that the enthalpies in the two processes coincide outside the transition region.
Thus, the specific heat $ c_p = \partial h / \partial T $ is readily obtained from Eq.~(\ref{enthalpy}), giving
\begin{equation}
c_p =
\begin{cases}
c_{ p,i }, & T \leq T_c \\
\begin{aligned}
& c_{ p,i } + ( c_{ p,m }-c_{ p,i } ) f \\ 
& + ( h_m - h_i ) \frac{ df }{ dT }, 
\end{aligned}
&  T_c  < T < T_c + \Delta \\
c_{ p,m }, & T \geq T_c + \Delta
\end{cases} .
\label{specific_heat}
\end{equation} 
In Fig.~\ref{fig5}, we show the specific heat and the enthalpy of the gate as a function of its temperature.

\begin{figure}
\includegraphics[width=\columnwidth]{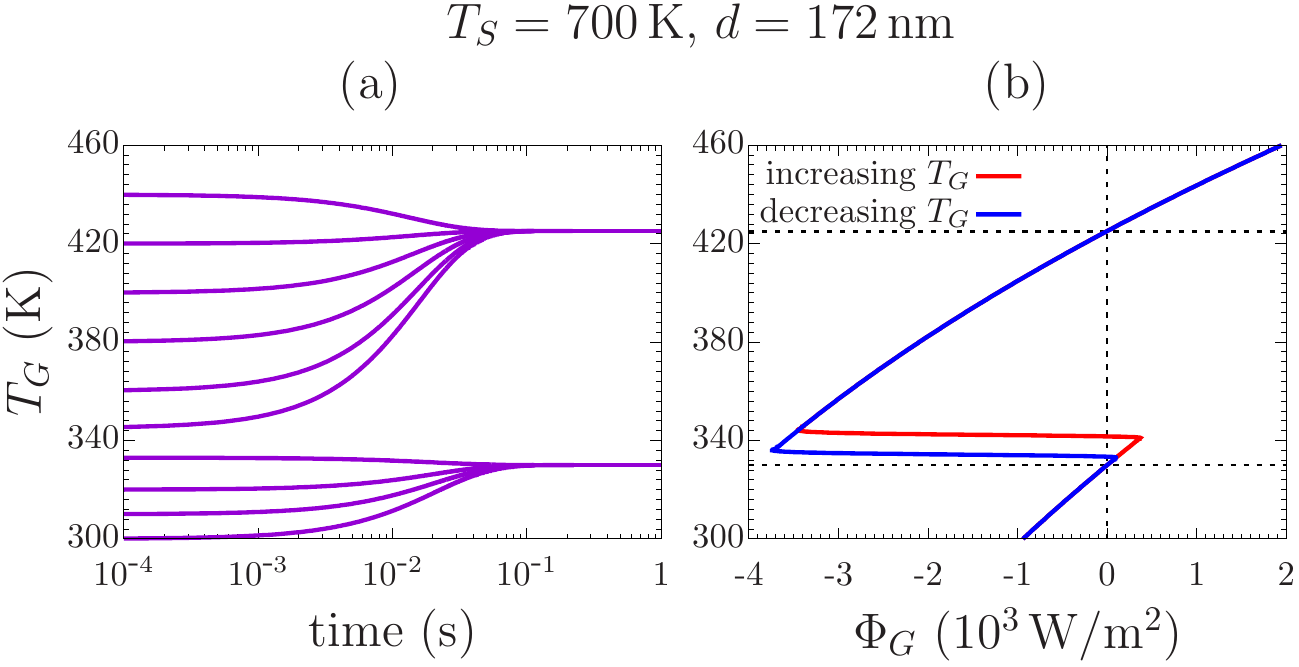}
\vspace{3mm}

\includegraphics[width=\columnwidth]{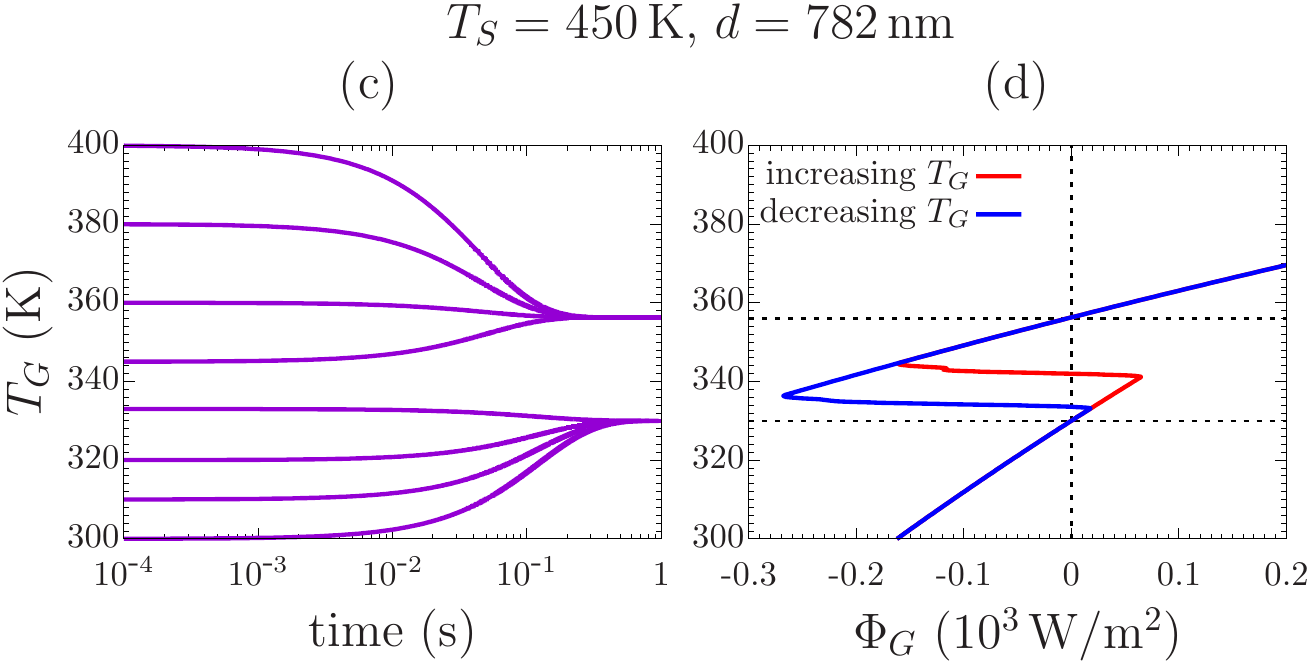}
\caption{Temporal evolution of the gate temperature at zero supplied energy flux ($\Phi_G=0$) for different initial temperatures. The temperatures $ T_G = 330\, $K and $ T_G = 425\, $K in (a) and $ T_G = 330\, $K and $ T_G = 356\, $K in (c) are stable stationary points. In (b) and (d) we show the corresponding $\Phi_G$ in the stationary regime.}
\label{fig6}
\end{figure}

\begin{figure}
\includegraphics[width=\columnwidth]{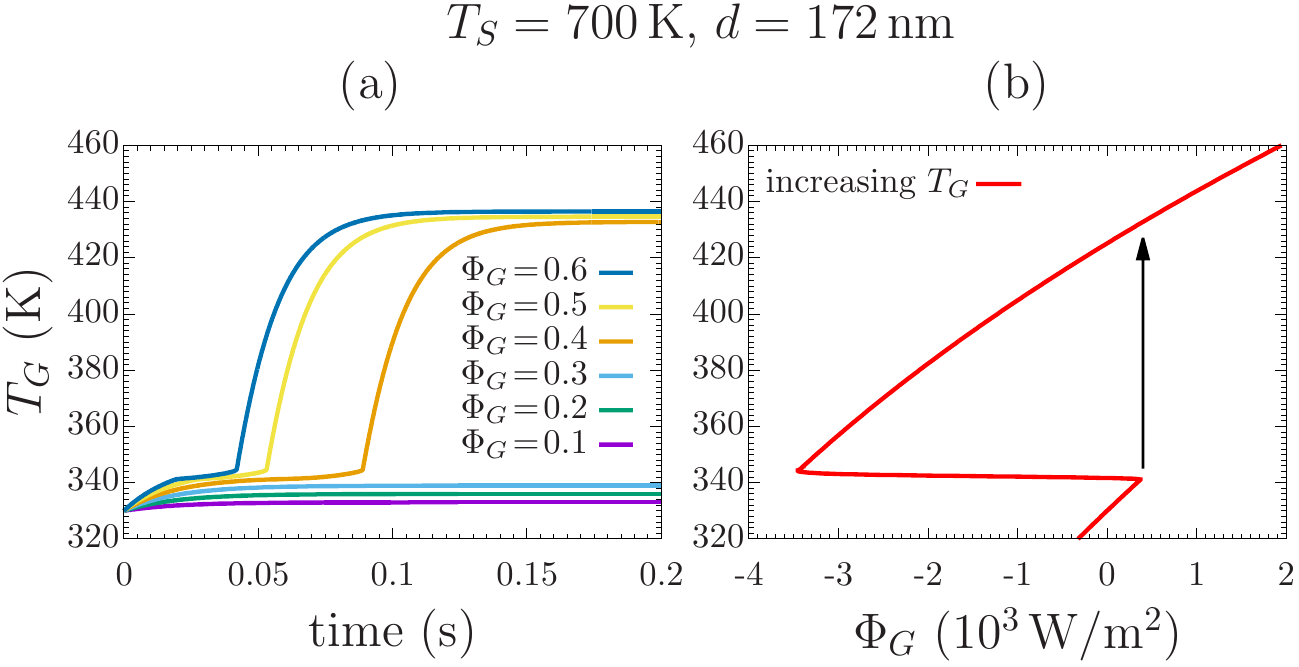}
\vspace{3mm}

\includegraphics[width=\columnwidth]{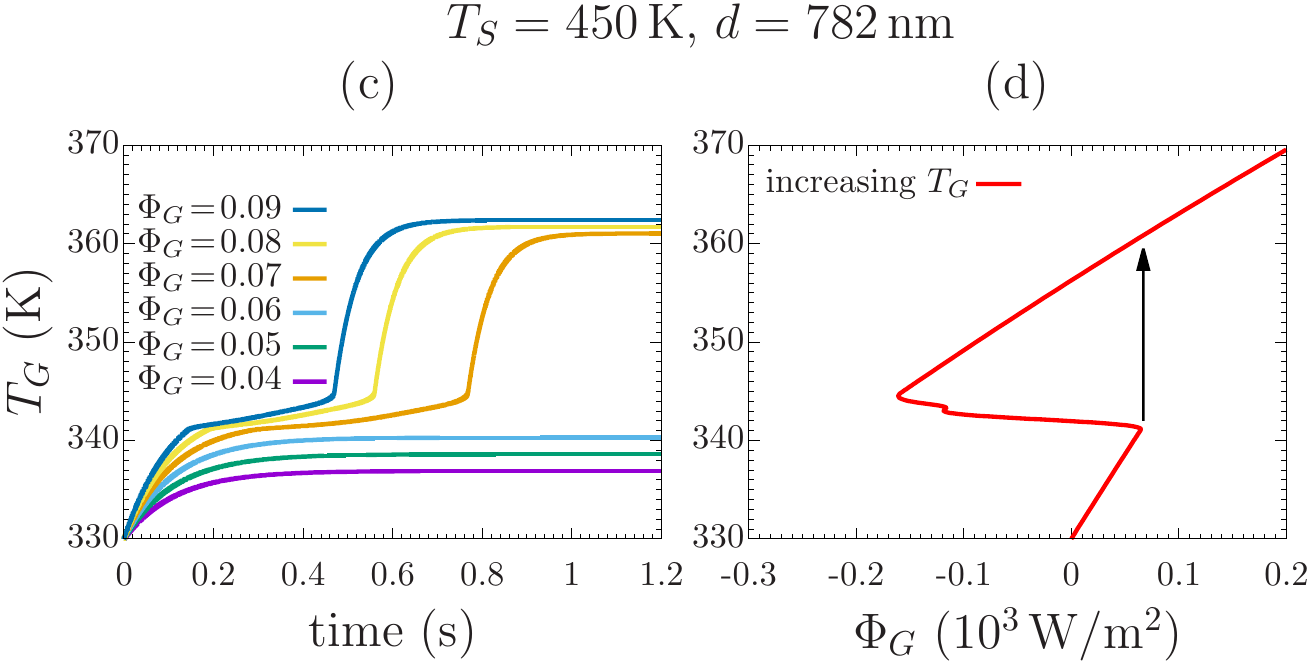}
\caption{Temporal evolution of the gate temperature $ T_G $ for an applied energy flux $ \Phi_G $ (in units of $10^3\:$W/m$^2$) and an initial temperature $ T_G(0) = 330\, $K. The applied fluxes $ \Phi_G $ in the stationary regime are also shown in (b) and (d) for the corresponding temperature $ T_G $.}
\label{fig7}
\end{figure}

Using the above description of the specific heat, the thermal inertia $ I_G $ can be determined as a function of $ T_G $ to compute the evolution of the temperature with Eq.~(\ref{evolution_temperature}). We first consider several initial temperatures $ T_G (0) $ that can be set by choosing the appropriate $ \Phi_G $ and then letting $ T_G (t) $ relax at $ \Phi_G = 0 $ towards the stationary regime. The results are shown in Fig.~\ref{fig6}(a) for the configuration with $ T_S = 700\, $K and $ d = 172\, $nm. We observe that for high enough initial temperatures, the gate evolves to the stationary state obtained in Sec.~\ref{sec:stationary_regime} by solving Eq.~(\ref{condition_eqtemp}), namely, at $ T_G = 425\, $K. For lower initial temperatures, in the stationary regime the gate approaches the equilibrium temperature we have imposed from the beginning, $ T_G = 330\, $K. 
The critical value $ T^\mathrm{ unst }_G $ for which $ T_G ( 0 ) < T^\mathrm{ unst }_G $ will lead to an evolution towards $ T_G = 330\, $K is located in the transition region. 
This behavior can be better understood from Fig.~\ref{fig6}(b), where we show $ \Phi_G $ in the stationary regime (horizontal axis) as a function of $ T_G $ (vertical axis). There, we observe that the stable equilibrium temperatures correspond to temperatures $ T_G $ such that $ \Phi_G = 0 $ with $ \Phi_G'>0$, while $ \Phi_G = 0 $ and $ \Phi_G'<0$ at the unstable point $ T^\mathrm{ unst }_G $. We emphasize that, because of the hysteresis, $ T^\mathrm{ unst }_G $ depends on how the membrane is treated.
In Fig.~\ref{fig6}(c), we show the evolution of $ T_G ( t ) $ in the configuration in which $ T_S = 450\, $K and $ d = 782\, $nm. As in the previous case, here there are also two stable stationary temperatures, $ T_G = 330\, $K and $T_G = 356\, $K. The corresponding flux $ \Phi_G $ as a function of $ T_G $ is shown in Fig.~\ref{fig6}(d).
Notice that since the gate has two stable states, the device can also work as a radiative thermal memory~\cite{Kubytskyi,ItoEtAl2016}. 

\begin{figure*}
\includegraphics[scale=1.17]{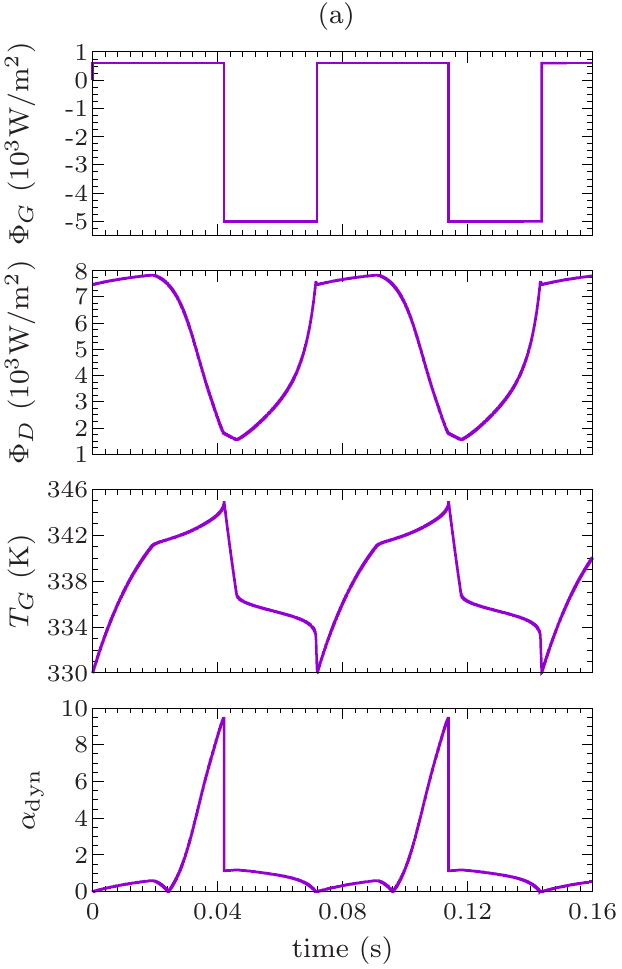}~\includegraphics[scale=1.17]{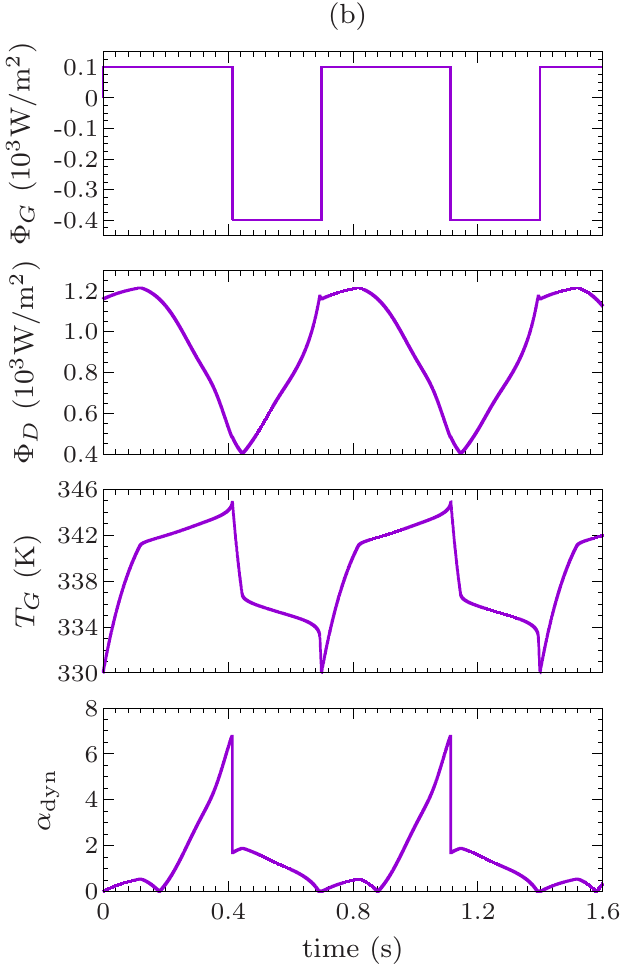}
\caption{Dynamical modulation of the radiative heat flux received by the drain. From top to bottom: Applied energy flux on the gate $ \Phi_G $, flux received by the drain $ \Phi_D $, temperature of the gate $ T_G $, and dynamical amplification factor $ \alpha_\mathrm{ dyn } $. In (a) with source temperature $ T_S = 700\, $K and separation distance $ d = 172\,$nm. In (b) with $ T_S = 450\, $K and $ d = 782\, $nm. The temperature of the drain is $ T_D = 300\, $K in both cases.}
\label{fig8}
\end{figure*}

Next we consider that in the initial state, the gate is in equilibrium at temperature $ T_G( 0 ) = 330\, $K, for which the corresponding applied flux is $ \Phi_G = 0 $. Then, a finite flux $ \Phi_G $ is applied that drives the evolution of the gate towards a new steady state. The resulting evolution is shown in Fig.~\ref{fig7}(a) for several values of $ \Phi_G $ for the configuration with $ T_S = 700\, $K and $ d = 172\, $nm. The corresponding results for the configuration with $ T_S = 450\, $K and $ d = 782\, $nm are shown in Fig.~\ref{fig7}(c). In these evolution curves, we observe the influence of the behavior of the specific heat in the transition region, which clearly slows down the dynamics of the gate. Notice that a band of forbidden temperatures is formed in the final steady state, which correspond to the temperature jumps pointed out in Refs.~\cite{Prodhome1,Ordonez} for far-field transistors. The appearance of a set of forbidden temperatures in the stationary regime is due to the mismatch in the optical properties of the two VO$_2$ phases. It can also be understood by inspection of the behavior of $ \Phi_G $ as a function of $ T_G $ in this regime.
The fluxes $ \Phi_G $ in the stationary regime (horizontal axis) as a function of $ T_G $ (vertical axis) corresponding to Figs.~\ref{fig7}(a) and~\ref{fig7}(c) are shown in Figs.~\ref{fig7}(b) and~\ref{fig7}(d), respectively. There, a jump to higher temperatures is indicated with an arrow at the local maximum of $ \Phi_G $ (seen as a function of $ T_G $) which coincides with the region of forbidden temperatures.

Starting at some stationary state, we now want to perform a dynamical modulation of the energy flux received by the drain by applying finite increments of energy flux on the gate. The temperatures of the source and the drain are assumed to be constant during the modulation. For a sequence of times $ \{ t_n \} $, $ n = 0,1,2\dots $, we consider the applied flux to be
\begin{equation}
\Phi_G( t ) =
\begin{cases}
\Phi_0, & t = t_0 \\
\Phi_h, & t_0 < t \leq t_1 \\
\Phi_c, & t_1 < t < t_2
\end{cases},
\label{applied_flux}
\end{equation}
and then let the modulation be periodic with period $ \tau = t_2 -t_0 $ by taking $ \Phi_G ( t + \tau ) = \Phi_G( t ) $ for all $ t $. Here $ \Phi_0 $ is the energy flux on the gate corresponding to the initial stationary state, $ \Phi_h > 0 $ is an energy flux applied to heat up the gate and $ \Phi_c < 0 $ is used to cool it down. Furthermore, we choose the modulation to start at $ t_0 = 0 $ in the state at which the gate is in equilibrium in the insulating phase, at $ T_G = 330\, $K, so that $ \Phi_0 = 0 $, and at time $ t = \tau $ the system comes back to the same equilibrium state. Since the temperature of the gate is periodically modulated by the action of $ \Phi_G $, the flux on the drain $ \Phi_D $ will be periodically modulated as well.

In order to quantify the performance of the modulation, we introduce a dynamical amplification factor 
\begin{equation}
\alpha_\mathrm{ dyn }( t ) = \left| \frac{ \Delta \Phi_D( t ) }{ \Delta \Phi_G( t ) } \right|,
\end{equation}
where $ \Delta \Phi_j ( t ) = \Phi_j ( t ) - \Phi_j ( t_0 ) $ for $ j = D,G $. Defined in this way, the amplification factor $ \alpha_\mathrm{ dyn } ( t ) $ accounts for the relative variation in time of $ \Phi_D $ with respect to the finite variation of $ \Phi_G $. 

Let us consider first the configuration with $ T_S = 700\, $K and $ d = 172\, $nm. As we said before, the modulation starts at $ t = 0 $ with the VO$_2$ membrane in the insulator phase. In the first step of the modulation, we applied a heat flux $ \Phi_h = 600\,$W/m$^2$ during a time $t_1$ such that the temperature of the gate rises from $ T_G ( 0 ) = 330\, $K to $ T_G ( t_1 ) = 345\, $K. The latter temperature correspond to the end of the transition region when $ T_G $ is increased, so that VO$_2$ is in the metallic phase at this point. Then, a heat flux $ \Phi_c = -5000\,$W/m$^2$ is applied until $ t = t_2 $ for which the temperature of the gate decreases to $ T_G ( t_2 ) = 330\, $K. The gate comes back to the initial state at this point and then we repeat the process. In Fig.~\ref{fig8}(a), we show for this case the input flux $ \Phi_G $, the modulated flux on the drain $ \Phi_D $, temperature of the gate $ T_G $, and the dynamical amplification factor $ \alpha_\mathrm{ dyn } $. In the configuration with $ T_S = 450\, $K and $ d = 782\, $nm, we implement the same protocol with $ \Phi_h = 100\,$W/m$^2$ and $ \Phi_c = -400\,$W/m$^2$. The results for this case are shown in Fig.~\ref{fig8}(b).

The previous procedure is an example of an active modulation of $ \Phi_D $ which illustrates the time scales of the device in operating conditions. In the first of the considered configurations ($ T_S = 700\, $K and $ d = 172\, $nm) the period is $ \tau \sim 0.07\, $s and the dynamics is about one order of magnitude faster than in the second case ($ T_S = 450\, $K and $ d = 782\, $nm). This difference in the dynamics is expected, since higher source temperatures and shorter separations distances induce stronger interactions of the gate with the source and the drain, respectively. Furthermore, we highlight that $ \alpha_\mathrm{ dyn } ( t ) $ is larger than unity in some intervals of the modulation, so that variations in time of $ \Phi_G $ can actually induce relatively larger variations of the modulated flux $ \Phi_D $.
It is worth mentioning that the time response of the transistor strongly depends on the thermal inertia of the gate and that the modulation rate is limited by the applied power driving the dynamics. 

\section{Summary and conclusions}

In this work, we have introduced a radiative thermal transistor made with a phase-change material based membrane (gate) in interaction in the near field with a substrate (drain) and in the far field with a thermal bath (source). We have shown that this transistor can operate at several time scales, depending on the temperature of the source, which are typically smaller than that in a radiative thermal transistor working in the far-field regime. 

We have also shown that the separation distance $ d $ between the gate and the drain can be chosen, for a given temperature of the source, to enforce a suitable equilibrium temperature for the gate. In our examples, for a drain at temperature $ T_D = 300\, $K, we have considered source temperatures $ T_S = 700\, $K and $ T_S = 450\, $K which for $ d = 172\, $nm and $ d = 782\, $nm, respectively, lead to a gate equilibrium temperature $ T_G^\mathrm{ eq } = 330\, $K at zero applied energy flux on the gate. This equilibrium temperature occurs below the transition region of the phase-change material (VO$_2$), and in these two examples there exists also a second equilibrium temperature above the transition region. By considering an applied flux on the gate periodically modulated, we have implemented a dynamical modulation of the energy flux received by the drain. In the case with higher source temperature and shorter separation distance, the dynamical response of the device is typically faster, since energy fluxes involved in this configuration are larger than those for the case with lower source temperature and larger separation distance. Furthermore, because of the negative differential thermal resistance induced by the phase-change material in the transition region, we have shown that the device can exhibit dynamical amplification factors larger than unity.

We emphasize here that energy losses in the gate due to conduction have not been included in the description of the device. Since the transistor is designed to manage radiative energy fluxes, conductive contributions have to be as small as possible. If these contributions were not negligible, however, one could compensate them by tuning the applied energy flux on the gate. On the other hand, energy losses by conduction could be used to cool down the gate. 

The device we have considered here could be employed to actively control the heat radiated to a medium by changing the temperature of the membrane around its critical value. Fast dynamical thermal management of nanostructures using non-contact devices could also be useful to develop microelectromechanical machines where heat is used to move microscopic devices, such as cantilevers, and also in microbolometer technologies to reduce the delay between two successive detections.

\begin{acknowledgments}

This work was partially supported by the Natural Sciences and Engineering Research Council of Canada (NSERC), Strategic Partnership Grants for Project program.

\end{acknowledgments}

\end{document}